\documentclass{svjour3}                     
\smartqed  
\usepackage{graphicx}
\journalname{Few-Body Systems (FB20)}
\begin{document}

\title{
On the proliferation of X's, Y's and Z's candidates
\thanks{Presented at the 20th International IUPAP Conference on Few-Body Problems in Physics, 20 - 25 August, 2012, Fukuoka, Japan}
}
\subtitle{
}


\author{A. Valcarce         \and
        T.F. Caram\'es      \and
        J. Vijande 
}


\institute{A. Valcarce \at
              Dpto. de F\'\i sica Fundamental, Universidad de Salamanca (Spain) \\
              \email{valcarce@usal.es}           
           \and
           T.F. Caram\'es \at
              Dpto. de F\'\i sica Fundamental, Universidad de Salamanca (Spain) \\
              \email{carames@usal.es}
           \and
           J. Vijande \at
              Dpto. de F\'\i sica At\'omica, Molecular y Nuclear, Universidad de Valencia e IFIC (Spain) \\
              \email{javier.vijande@uv.es}
}

\date{Received: date / Accepted: date}

\maketitle

\begin{abstract}
Heavy meson spectroscopy above open flavor thresholds has become a challenge both from the
experimental and theoretical points of view. Experimentally, several signals have been
interpreted as meson resonances with unusual properties; theoretically, such signals may be
identified with meson-meson molecules or compact multiquark structures. We analyze the 
influence of thresholds on heavy meson spectroscopy comparing different
flavor sectors and quantum numbers. The validity of a quark-model picture above
open-flavor thresholds would severely restrict the number of channels that may lodge 
multiquark structures as meson-meson molecules.  

\end{abstract}
\vspace*{0.5cm}

The discoveries in 1974 of the so-called {\it November revolution}~\cite{Nov74},
as the name 'revolution' implies, were not just additions to our knowledge
of nature. Instead they signalled a change in our understanding of the 
structure of matter. Of course, this change did not occur completely
overnight. Many of the ideas were there before, accepted by some and doubted by 
others. As always, there were also plenty of wrong ideas and irrelevant pieces
of information. Why were these discoveries so exciting and so significant?
First, because those particles were made of an unknown charm quark, and second
because the spectrum of such particles could be studied with great precision and
it was just the spectrum one would expect on the basis of quark constituents~\cite{Teo75}.
We had a stunning example which showed the quark layer of substructure to matter~\cite{Gil84}.

Since 1974 to 2003 all states discovered on charmonium spectroscopy fitted nicely into the
simple spectroscopic models based on the one-gluon exchange postulated in 1975~\cite{Bar05}. The
same theoretical ideas were applied to mesons made of a light and a heavy quark with great 
success~\cite{God85}. In 2003 we relived a kind of revolution with the discovery
by BABAR~\cite{Bar03} of an open-charm meson, the $D_{s0}^*(2317)$, whose mass seemed to contradict
current spectroscopic models; and by Belle~\cite{Bel03} of a charmonium-like state, the $X(3872)$,
with intriguing properties to be a simple quark-antiquark pair. 
Since then, the closely tended garden of heavy mesons is
abloom with exotic new growths. These discoveries are offering exciting new
insights into the subtleties of the strong interaction, awaiting for a 
general explanation that could lead
to a more unified and mature picture of hadron spectroscopy. A recurring old question
arises again: being some of these new states firmly established (as it is the case of 
the $X(3872)$), could they be fitted into the quark
model scheme or are we in front of the breakdown of such a pattern?

The question above has been the flagship of many experimental and theoretical
efforts during the last two decades~\cite{Isg83}. It was already in May 1993, being N. Isgur the 
Cebaf theory group leader, when he invited the few students participating at the 
Hugs@Cebaf summer-school for lunch. He confessed that
what he would like more to learn from Cebaf was the reason why the constituent quark model
was wrong. Some of the students had to write a proceeding paper based on Isgur's
lectures, it was entitled {\it Before the breakdown of the non-relativistic 
quark-model}~\cite{Bal93}. We had the chance to talk to Isgur a few years later, 
in the 1996 Confinement conference. He laughed remembering that anecdote
and he said {\it I keep trying}. In fact, one of his last papers dealt with the
impact of thresholds on the hadronic spectrum, in an attempt to go beyond the
adiabatic approximation within the constituent quark model~\cite{Isg99}. 

Ten years later, F. Close entitled 
the summary talk at Hadron03~\cite{Clo03} {\it The end of the
constituent quark model?} These were the hectic times of the $\theta^+$ affair.
He also pointed out that listing all of the mesons
from the PDG as a function of $J^{PC}$ indicated that the light hadron
dynamics is clearly overpopulated, showing that not all of the data could
be correct. The question of when does the constituent quark model work was once again
made, noting that its successful picture for charmonium gets significant distortion
from the $D\bar D$ threshold region. Thus, the relevance of thresholds
was posed as an important potential distortion for the 
predictions of the constituent quark model. 

Two final comments on this historical perspective. C. Quigg in the theory summary talk
at Hadron11~\cite{Qig11}, {\it The future of hadrons}, put some caution
about the proliferation of low-statistics experimental data, pointing out that
an experimental signal of $3\sigma$ means that it will probably
dissapear 80\% of times. Finally, the difficulty of disentangling resonances from
cusps due to the opening of thresholds has been the matter of study to give an
alternative explanation to some of the recently reported new states~\cite{Bug11}.

Our purpose in this talk is to discuss how the new set of states reported could offer
insight to check the validity of the constituent quark model beyond flavor thresholds.
It would severely restrict the number of channels that may lodge meson-meson
molecules. Out of the many states recently reported in charmonium and bottomonium 
spectroscopy, we do not really know how many will survive future experimental 
screenings. We are still shocked by the $\theta^+$ resonance, seen by so many experiments 
that later on did not find anything~\cite{The12}. There is only one state that has been
firmly established by different collaborations and whose properties seem 
to be hardly accommodated in the quark-antiquark scheme, this is the $X(3872)$~\cite{Bel03}.
Regarding the zoo of other states that have been proposed, we have to stay tuned
but also be cautious~\cite{Clo03,Qig11}. Some members of this $XYZ$ jungle are awaiting confirmation,
seen only by one collaboration, like the intriguing charged 
state $Z(4430)$, seen by Belle but not by BABAR~\cite{Cho08}. Other members of this jungle 
cannot be excluded to fit into the simple quark-antiquark scheme, like the
$Z(3930)$ recently identified as the $\chi_{c2}(2P)$ charmonium state~\cite{Aub10}. 
Other experimental signals seen only by a particular experiment, in some of the 
expected decay modes, or with low statistics, could just be the reflection of the opening
of thresholds~\cite{Bug11}. 
\begin{figure*}
\includegraphics[width=12cm,height=17cm]{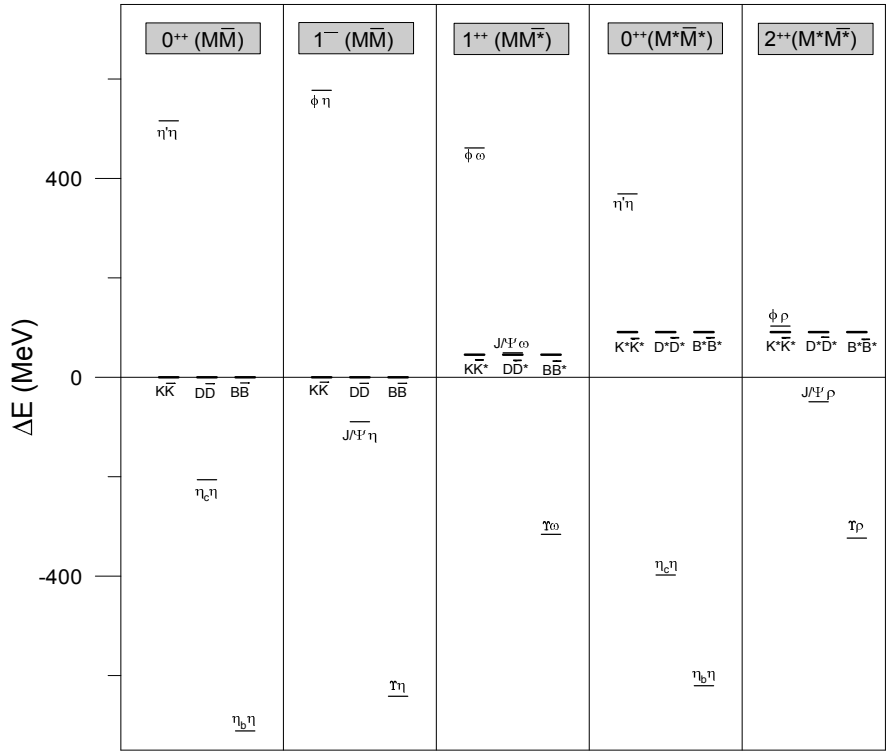}
\vspace*{-6.5cm}
\caption{Experimental thresholds of four-quark systems made of a heavy and a light
quark and their corresponding antiquarks, $Qn\bar Q \bar n$ with 
$Q=s$, $c$, or $b$, for several sets of quantum numbers, $J^{PC}$. We have set as 
our origin of energies the $K\bar K$, $D\bar D$ and $B\bar B$ masses for the hidden
strange, charm and bottom sectors, respectively.}
\label{fig1}
\end{figure*}

Let us start by discussing Fig.~\ref{fig1}. In this figure we have plotted 
the experimental thresholds of four-quark systems made of a heavy and a light
quark and their corresponding antiquarks for several sets of quantum numbers, $J^{PC}$, in 
three different flavor sectors: $Q=s$, hidden strange; $Q=c$, hidden charm; and 
$Q=b$, hidden bottom. In every flavor sector we represent the mass difference 
with respect to the mass of $K\bar K$,
$D\bar D$ and $B\bar B$, respectively. In a constituent quark model picture, the four-quark state
$Qn \bar Q \bar n$ could either split into $(Q \bar n) - (n \bar Q)$ or
$(Q \bar Q) - (n \bar n)$. One observes how the general trend for all quantum 
numbers is that the mass of the $(Q \bar Q) - (n \bar n)$ two-meson system is 
larger than the mass of the $(Q \bar n) - (n \bar Q)$ two-meson state for $Q=s$, but it is 
smaller for $Q=c$ or $b$. It is remarkable the case
of $J^{PC}=1^{++}$ for $Q=c$, where the $(Q\bar Q) - (n\bar n)$ and 
the $(Q \bar n) - (n \bar Q)$ two-meson states are almost degenerate. The reverse of the ordering of
the masses of the $(Q \bar Q) - (n \bar n)$ and $(Q \bar n) - (n \bar Q)$
thresholds when increasing the mass of the heavy quark for all $J^{PC}$
quantum numbers can be simply understood within the constituent quark model
with a Cornell-like potential~\cite{Clo03}. 
The binding of a coulombic system is proportional to the reduced mass of the
interacting particles. Thus, for a two-meson threshold with a heavy-light light-heavy 
quark structure, the binding of any of the two mesons does not change much when increasing the mass 
of the heavy flavor, due to the reduced mass of each meson being close
to the mass of the light quark. However, if the two-meson state presents a
heavy-heavy light-light quark structure, the binding of the heavy-heavy meson increases with the mass of 
the heavy particle while that of the light-light meson remains constant,
becoming this threshold lighter than the heavy-light light-heavy two-meson structure,
as seen in Fig.~\ref{fig1}. 

Such a picture, together with the absence of long-range
forces~\cite{Tor92} in a charmonium-light two-meson system,
may imply different consequences for the existence of molecules
close to the meson-antimeson threshold.
First, the possible existence of such molecules in the 
hidden-strange sector. If the $K\bar K$ interaction is attractive for some particular set
of quantum numbers, this two-meson system may be stable because 
no other threshold appears below, the dissociation of the molecule
being therefore forbidden (see the $Q=s$ states for
any $J^{PC}$ quantum numbers in Fig.~\ref{fig1}). Such a possibility would become
more probable for those quantum numbers where the quark model
seems to work worst, those cases where one needs a P-wave in the
simplest quark model structure, $q \bar q$, but can be obtained in
S-wave from a four-quark system. In these cases, the mass of the four-quark system
could be even below the predicted lowest quark-antiquark state.
This is precisely the idea suggested by Weinstein and
Isgur~\cite{Wei90} as a plausible explanation of the proliferation of 
scalar mesons in the light quark sector. They concluded the $J^{PC}=0^{++}$ and $1^{++}$
quantum numbers to be the best candidates to lodge a meson-antimeson molecule.
These quantum numbers are P-wave in the quark model but S-wave in the four-quark picture
and besides they are spin triplet, having therefore an attractive spin-spin 
interaction. 

Second, the possibility of finding meson-antimeson molecules contributing to 
the meson spectrum becomes more and more difficult when increasing the mass
of the heavy flavor, due to the lowering of the mass of the $(Q \bar Q) - (n\bar n)$ 
threshold (see the $Q=c$ or $b$ states for any $J^{PC}$ quantum numbers in Fig.~\ref{fig1}).
This would make the system dissociate immediately. In such cases,
the presence of attractive meson-antimeson thresholds would manifest in the scattering cross section but 
they will not lodge a physical resonance. These ideas favored 
the interpretation of several of the experimental
signals in charmonium and bottomonium spectroscopy above flavor thresholds as originated from the 
opening of the threshold and not being resonances~\cite{Bug11}. 

Thus, only a few channels may lodge molecular resonances. As discussed above, there is a remarkable exception to the general rule,
the $(I)J^{PC}=(0)1^{++}$ quantum numbers in the charmonium sector. In this case the 
$(c \bar n) - (n\bar c)$ ($D\bar D^*$) and $(c\bar c) - (n\bar n)$ ($J/\Psi \omega$) thresholds
are almost degenerate, and the attractive $D\bar D^*$ interaction together
with the cooperative effect of the almost degenerate two-meson thresholds
give rise to the widely discussed $X(3872)$~\cite{Car09}. 
In spite of the general idea that the stability of a system made of quarks
comes favored by increasing the mass of the heavy flavor, it becomes more complicated when several vectors in color
space contribute to generate a color singlet, as it is the case of four-quark systems~\cite{VijPR}. 
The reason is that, as explained above, the mass of one of the thresholds,
$(Q\bar Q) - (n\bar n)$, diminishes rapidly when the heavy quark mass increases (see Fig.~\ref{fig1}), making
therefore the meson-antimeson system, $(Q\bar n) - (n\bar Q)$, unstable. This simple reasoning
of coupled-channel calculations was illustrated in Ref.~\cite{Car12}.
It is the coupling to the almost degenerate $J/\Psi \omega$ 
channel the responsible for having a bound state just
below the $D\bar D^*$ threshold. Such an explanation
comes reinforced by the recent observation of the decay 
$X(3872) \to J/\Psi \omega$~\cite{Amo10}. When the mass of the heavy
quark is augmented from charm to bottom, the $B \bar B^*$ becomes more
attractive due to the decreasing of the kinetic energy and having
essentially the same interaction. However, the coupling to the lower channel,
$\Upsilon \omega$, destroys any possibility of having a bound state. Thus, 
based on the constituent quark model ideas, one should not expect
a twin of the $X(3872)$ in bottomonium spectroscopy like those pointed out 
in hadronic models based on the traditional meson theory of the nuclear forces 
or resorting to heavy quark symmetry arguments~\cite{Tor92}.
Finally, the coupling to channels containing the light pion destroys the degeneracy between meson-antimeson
and charmonium-light two-meson thresholds, an important mechanism for binding four-quark states
in the $I=0$ sector. This excludes, for example, the existence of charged partners of the
$X(3872)$, as explained in Ref.~\cite{Car09}. Only one S-wave channel, the $J^{PC}=2^{++}$, where the
coupling to the charmonium-pion two-meson system is prohibited, may be candidate for lodging
a resonance close above the $D^* \bar D^*$.

The proliferation of resonances above flavor thresholds 
could rely on our poor knowledge of confinement. 
Refs.~\cite{Vij12} have analyzed the stability
of $Q n \bar Q \bar n$ and $Q Q\bar n \bar n$ systems by
considering only a multiquark confining interaction in an attempt
to discern whether confining interactions not factorizable as
two-body potentials would influence the stability of four-quark
states. The ground state of systems made of two quarks and
two antiquarks of equal masses was found to be below the
dissociation threshold. Whereas for the cryptoexotic $Q n\bar Q \bar n$
the binding decreases with increasing mass ratio $m_Q/m_n$,
for the flavor exotic $Q Q \bar n \bar n$
the effect of mass symmetry breaking is opposite. 

The discussion on the last paragraph drives us to a brief comment on exotic states
$QQ \bar n \bar n$. In this case the situation is rather different to the nonexotic
$Qn \bar Q \bar n$ system, because the possible
dissociation thresholds do not contain
states made of a heavy quark and a heavy antiquark, whose binding would increase linearly
with the mass of the heavy flavor.
Thus, stability will be favored by increasing the mass of the heavy
flavor, being much more probable in the bottom sector than in the
strange one~\cite{Car11}. The search of such exotic
states is a hot experimental subject for the incoming years in different experimental
facilities~\cite{Exp11}.

Summarizing, recent experimental data on charmonium spectroscopy have suggested the
existence of a large number of states above charmed meson thresholds.
They have been baptized as $X's$, $Y's$, and $Z's$, due to their unusual properties
not easily explained in terms of simple quark-antiquark pairs. Such proliferation of states
has pointed out to the existence of meson-antimeson molecules. In a quark-model picture we have justified
how such molecules may contribute to the light meson spectroscopy. In particular,
they could explain the existence of non quark-antiquark states for quantum numbers
that can be obtained from four-quarks in an S-wave but need orbital angular momentum from
a quark-antiquark pair. When increasing the mass of the heavy flavor, the possibility of
having meson-antimeson resonances decreases with the mass of the heavy quark. Only in some
particular cases the cooperative effect of nearby two-meson channels with an attractive
meson-antimeson interaction may produce resonances in the charmonium sector, the $X(3872)$ being the example
par excellence. Increasing the mass of the heavy quark destroys the possibility of a
twin state in the bottom sector, against the predictions of hadronic models based on the 
traditional meson theory of the nuclear forces or heavy quark
symmetry. Improved confinement interactions considering many-body forces go against
the stability of non-exotic four-quark states
in the energy region close to the flavor thresholds.
Finally, in the exotic sector, due 
to the nonexistence of thresholds made of two heavy quarks, the stability of two-meson states
would increase with the mass of the heavy quark.

\begin{acknowledgements}
This work has been partially funded by the Spanish Ministerio de
Educaci\'on y Ciencia and EU FEDER under Contract No. FPA2010-21750,
and by the Consolider-Ingenio 2010 Program CPAN (CSD2007-00042).
\end{acknowledgements}


\end{document}